\documentclass[twocolumn]{aastex631}

\usepackage{graphicx}
\usepackage{soul,xcolor}
\usepackage{amsmath}
\usepackage{multirow}
\usepackage{hyperref}

\graphicspath{{figures/}}  

\usepackage{natbib}
\usepackage{CJKutf8}
\newcommand{\cntext}[1]{\begin{CJK*}{UTF8}{bsmi}#1\end{CJK*}}

\revised{\today}
\submitjournal{ApJ}

\shorttitle{Companion of SN 2020oi}
\shortauthors{Chen, Rau, \& Pan}

\begin{document}

\title{Exploring the Observability of Surviving Companions of Stripped-Envelope Supernovae: A Case Study of Type Ic SN 2020oi}

\newcommand*{\NTHUP}{Department of Physics, National Tsing Hua University, Hsinchu 30013, Taiwan}
\newcommand*{\NTHUA}{Institute of Astronomy, National Tsing Hua University, Hsinchu 30013, Taiwan}
\newcommand*{\CICA}{Center for Informatics and Computation in Astronomy, National Tsing Hua University, Hsinchu 30013, Taiwan}
\newcommand*{\UIUC}{Department of Astronomy, University of Illinois at Urbana-Champaign, Urbana, IL 61801, USA}
\newcommand*{\CTC}{Center for Theory and Computation, National Tsing Hua University, Hsinchu 30013, Taiwan}
\newcommand*{\NCTS}{Physics Division, National Center for Theoretical Sciences, National Taiwan University, Taipei 10617, Taiwan}

\author[0000-0003-1640-9460]{Hsin-Pei Chen (\cntext{陳昕霈})}
\affiliation{\NTHUA} \affiliation{\NTHUP} \affiliation{\CICA}

\author[0000-0003-4692-5941]{Shiau-Jie Rau (\cntext{饒孝節})}
\affiliation{\UIUC}

\author[0000-0002-1473-9880]{Kuo-Chuan Pan (\cntext{潘國全})}
 \affiliation{\NTHUA} \affiliation{\NTHUP} \affiliation{\CICA} \affiliation{\CTC} \affiliation{\NCTS}

\begin{abstract}
Stripped-envelope supernovae (SE SNe) were considered as the explosions of single massive stars with strong stellar winds, while later observations favor binary origins. One direct evidence to support the binary origins is to find the surviving companions of SE SNe since previous numerical studies suggested that the binary companion should survive the supernova impact and could be detectable.
Recently, \cite{2022ApJ...924...55G} reported that the nearby Type~Ic SN~2020oi in M100 ($\sim$ 17.1 Mpc) resulted from a binary system based on the HST photometric and spectroscopic observation.
Based on the suggested binary properties of SN 2020oi, we conduct two-dimensional hydrodynamics simulations of supernova-companion interactions and the subsequent post-impact evolution of the companion. 
Our results suggest that a surviving companion becomes brighter in two orders of magnitude and temporarily redder after the SN impact. The companion might be detectable with the JWST NIRCam short wavelength channel in a few years. Furthermore, the predicted magnitudes of surviving companions show a significant magnitude gradient around the peak. This could be another indicator to identify the surviving companion from a SE SN.

\end{abstract}


\section{INTRODUCTION}
\label{sec:intro}

Stripped-envelope supernovae (SE~SNe) are core-collapse supernovae (CCSNe) with few or without hydrogen (H) and helium (He) lines because the progenitors' envelopes are partially or entirely stripped before the SN explosion \citep{2019NatAs...3..717M}.
Two mainstream scenarios to explain envelope stripping are the stellar-wind origin \citep{1975ApJ...195..157C} and the binary-interaction origin \citep{1975ApJ...200..145W}.
The stellar-wind scenario requires a massive progenitor ($\gtrsim$ 25 $\rm M_{\odot}$) that blows out its envelope via a strong stellar wind before the SN explosion \cite[e.g.][]{2014Natur.509..471G, 2014A&A...572L..11G}.
On the other hand, the binary-interaction scenario suggests that the SN progenitor star has been interacting with a binary companion and therefore lost its envelope via mass transfer \citep{2012ARA&A..50..107L}.
In the past decade, a few pieces of evidence from both theoretical and observational perspectives prefer the binary-interaction origin.
For instance, \cite{2012Sci...337..444S} spectroscopically analyzed the orbital parameters of nearby massive stars and concluded that more than 70\% of massive stars had experienced mass transfer during their stellar evolution. 
\cite{2012A&A...544L..11Y} discussed the possible progenitors of type Ib/c SNe by comparing the stellar evolution models of massive helium stars ($M=3-5 \rm M_{\odot}$) to the observations of galactic WR stars ($M>8 \rm M_{\odot}$). They concluded that 
the relatively low-mass helium stars with a thick helium envelope can be as bright in visual bands and the SN progenitor mass could be overestimated, which disfavor the massive single star origin with stellar wind.
In addition, \cite{2009MNRAS.395.1409S}, \cite{2013MNRAS.436..774E}, and \cite{2015PASA...32...16S}, analyzed nearby CCSNe with identified pre-explosion progenitors or luminosity upper limits. Connecting these observed populations with theoretical interpolations, they concluded that their data prefer low-mass progenitors stripped by their binary companions and high mass ($M_{\rm init} \gtrsim 18 \, \rm M_\odot$) SN progenitors are rare.
\cite{2019MNRAS.485.1559P} analyzed 18 SE SNe happened during 2013-2018 and found that most of these SE SNe have progenitor mass $< 25 \, \rm M_\odot$, and ejecta mass distribution shows only one population, implying that the binary interaction scenario dominates the origin of SE SNe.

One direct evidence to support the binary-interaction origin of SE SNe is to search for surviving companions in supernova remnants (SNRs).
SN-companion interaction is studied thoroughly in the simulations of Type Ia supernovae \citep{1981ApJ...243..994F, 2010ApJ...715...78P, 2012ApJ...750..151P, 2012ApJ...760...21P, 2013ApJ...773...49P, 2015MNRAS.454.1192L, 2019ApJ...887...68B, 2020ApJ...898...12Z, 2021MNRAS.500..301L, 2022ApJ...933...38R} in the single-degenerate scenario,  
saying a carbon-oxygen white dwarf interacting with a non-degenerate companion.
The overall results show that, during the supernova-companion interactions, the SN ejecta strips and ablates the companion envelope.
The SN-impacted companions will survive with manifest signatures thousands of years after the impact: much more luminous, inflated, reddening, surface contaminated, and kicked by the momentum transfer. Such outlook changes contribute to the observation of examining the theoretical hypotheses.

Although similar studies have been applied to CCSNe in binaries as well, unlike Type~Ia SNe, the binary configurations have a much wider diversity in CCSNe. For instance, \citet{2014ApJ...792...66H} focused on the SN impact in an evolved massive-star binary (each $\sim$10 $\rm M_{\odot}$). They found that the companion lost its mass up to 25\% in the smallest separation cases due to its loose envelope. \citet{2015A&A...584A..11L}, motivated by the origin of hypervelocity stars, focused on the main-sequence (MS) companion with smaller mass, 0.9 and 3.5 $\rm M_{\odot}$ in a Type  Ib/c  SN  system.  They found that the SN impact causes $< 5\%$ of mass loss on the MS companions, and only the closest separation models have the kick velocity of post-impact companion up to $\sim$100 km $\rm s^{-1}$ and can potentially induce hypervelocity stars.
In addition, \citet{2018ApJ...864..119H} conducted several ejecta-companion interaction simulations with a wide parameter space of both the SN and the companion. Due to the discrepancy in kick velocity of the companion (the “impact velocity” defined in their paper) between their results and \cite{1975ApJ...200..145W}, they proposed an alternative model regarding momentum transfer and energy injection efficiencies. They found the kick velocity and mass loss of the SN-impacted companion are small, and the companion luminosity showed 1-2 order increases, which is comparable to the results in Type Ia simulations. In their subsequent study, \citet{2021MNRAS.505.2485O} further explored the post-impact companion evolution and estimated a 1-3 \% probability of observing a surviving companion in SE~SN systems.

On the other hand, observations of surviving companions of CCSNe are rare but growing recently.
The very first direct evidence was the Type IIb SN 1993J, which has been identified as a massive binary with pre-SN mass transfer, and the companion was first detected by the Hubble Space Telescope (HST) 10 years after the explosion \citep{2004Natur.427..129M}.
iPTF13bvn was the first Type Ib SN with an identified pre-explosion progenitor star \citep{2013ApJ...775L...7C, 2014AJ....148...68B}. Both SED fitting and modeling results prefer the binary models and ruled out a WR progenitor. However, the companion mass should be much lower than the progenitor since it has not been detected in the late-time photometry data \citep{2016MNRAS.461L.117E}.
Recently, two surviving companion candidates of SE SNe were reported: First, the surviving companion of Type Ib/c SN 2013ge is considered as a supergiant crossing the Hertzsprung Gap based on the comparison of HST observation and theoretical evolutionary tracks \citep{2022ApJ...929L..15F}.
Second, the pre-SN observation of Type Ib SN 2019yvr suggests an inflated yellow hypergiant companion \citep{2022MNRAS.510.3701S}.
Note that both of the studies suggest a non-MS companion. However, the binary population study of SE SNe found that $\sim$ 70\% of their companions are in the MS stage \citep{2017ApJ...842..125Z}.

In this study, we conduct supernova-companion interaction simulations and the subsequent post-impact evolution of surviving companions.
We adopt the properties of SN 2020oi \citep{2022ApJ...924...55G} due to its proximity and probable binary origin. SN~2020oi is a Type~Ic SN first detected on January 7th, 2020, in the nearby spiral galaxy M100. \citet{2022ApJ...924...55G} suggest the binary-interaction origin because of the low derived ejecta mass.
Their binary evolution models in \texttt{BPASS} indicate a binary separation below $10^{12}$ cm and a MS companion with a radius below 2 $\times 10^{11}$ cm at the time of the SN explosion. The properties of this system applied in our simulations are summarized in Table~\ref{tab:SN2020oi}.
Based on these constraints, our simulations could apply to a system containing a low mass ($\rm < 10 M_{\odot}$) SE SN and a MS companion.

The paper is structured as the following. Section \ref{sec:methods} describes the numerical methods and setup for the explosion models as well as the companion models. The physical parameters of our simulations are summarized in Table \ref{tab:modelsummary}. In Section \ref{sec:results}, we first describe the supernova-companion interaction simulations, including the binary separation dependence and the explosion energy dependence, then we describe the post-impact evolution of the surviving companions. In Section \ref{sec:discussion}, we discuss the observability of the surviving companions in SN~2020oi-like systems and introduce the observation of surviving companion candidates of SE SNe. Furthermore, we compare the evolution of surviving companions using methods described by \citet{2021MNRAS.505.2485O} with our implementation. Finally, we summarize our results and draw conclusions in Section \ref{sec:sum&con}.

\begin{deluxetable}{ccccc}
\tablewidth{0pt}
\tablenum{1}
\label{tab:SN2020oi}
\tablecaption{Summary of the physical properties of SN 2020oi system \cite{2022ApJ...924...55G} applied in our simulations.}
\tablehead{Object\tablenotemark{*} & Property & Notation & \multicolumn{2}{c}{Value}}
\startdata
SN & explosion energy   & $E_{\rm SN}$ & $\rm 7.9\times10^{50}$  & erg \\
SN & explosion velocity & $v_{\rm SN}$ & $\rm 1.275 \times 10^9$ & cm/s \\
SN & ejecta mass        & $M_{\rm ej}$ & 0.81                    & $\rm M_{\odot}$\\
SN & metallicity        & Z            & 0.75                    & $\rm Z_{\odot}$ \\
BS & separation         & A            & $\rm < 10^{12}$         & cm \\
CP & radius & $\rm R_{*}$ & $\rm < 2\times 10^{11}$              & cm
\enddata
\tablenotetext{*}{SN = supernova; CP = companion; BS = binary system}
\end{deluxetable}

\section{NUMERICAL METHODS}
\label{sec:methods}

The methodology of our simulations is similar to what has been reported in \cite{2012ApJ...750..151P, 2012ApJ...760...21P, 2014ApJ...792...71P}, and \cite{2022ApJ...933...38R} in Type~Ia SN simulations, but we apply it to a Type~Ic SN system and other similar systems. Here, we briefly review the setup of our simulations for the completeness of the paper. 
The simulations are separated into three steps: Firstly, the companion models are generated by the 1D stellar evolution code \texttt{MESA} (Modules for Experiments in Stellar Astrophysics; \citealt{2011ApJS..192....3P, 2013ApJS..208....4P, 2015ApJS..220...15P, 2018ApJS..234...34P, 2019ApJS..243...10P}; version: r12115; mesasdk version: 200301).
Secondly, the \texttt{MESA} models are artificially interpolated into the 2D \texttt{FLASH}\footnote{https://flash.rochester.edu/site/} (version 4) framework to complete the supernova impact simulations; \texttt{FLASH} is a multi-dimensional Eulerian hydrodynamics simulation code with adaptive mesh refinement (AMR) technology \citep{2000ApJS..131..273F, 2008PhST..132a4046D}. Lastly, after the ejecta-companion interaction processes, the post-impact companion evolution simulations are done by \texttt{MESA} again.
The \texttt{MESA} simulation codes and input files are available on Zenodo under an open-source Creative Commons Attribution license: \dataset[doi: 10.5281/zenodo.7777107]{https://doi.org/10.5281/zenodo.7777107}.

\subsection{Binary models}
We generate the companion models referring to the binary system of SN 2020oi predicted by \citet{2022ApJ...924...55G} (see  Table~\ref{tab:SN2020oi}). From their \texttt{BPASS} binary evolution models, they suggest the constraints of the companion radius $R_* <2 \times 10^{11}$ cm and the binary separation $A < 10^{12}$ cm. By considering a low-mass companion within the zero-age main sequence (ZAMS) stage, we choose 3.0, 5.5, and 8.0 $\rm M_{\odot}$ as the companion mass in our models, where 8.0 $\rm M_{\odot}$ models slightly exceed the companion radius constraint for a wider parameter space.

Note that, in the paper, we assume the companion stars are still in the ZAMS stage when the SN explosion begins. In reality, these companion stars would have evolved for several million years before the primary star exploded. However, study of binary populations \citep{2017ApJ...842..125Z} suggests that many companions of SE~SNe are still in the main sequence stage. To simplify the problem and minimize the impact of structural changes during the main sequence phase, we have only considered stars in the ZAMS stage.

\begin{figure}
\epsscale{1.1}
\plotone{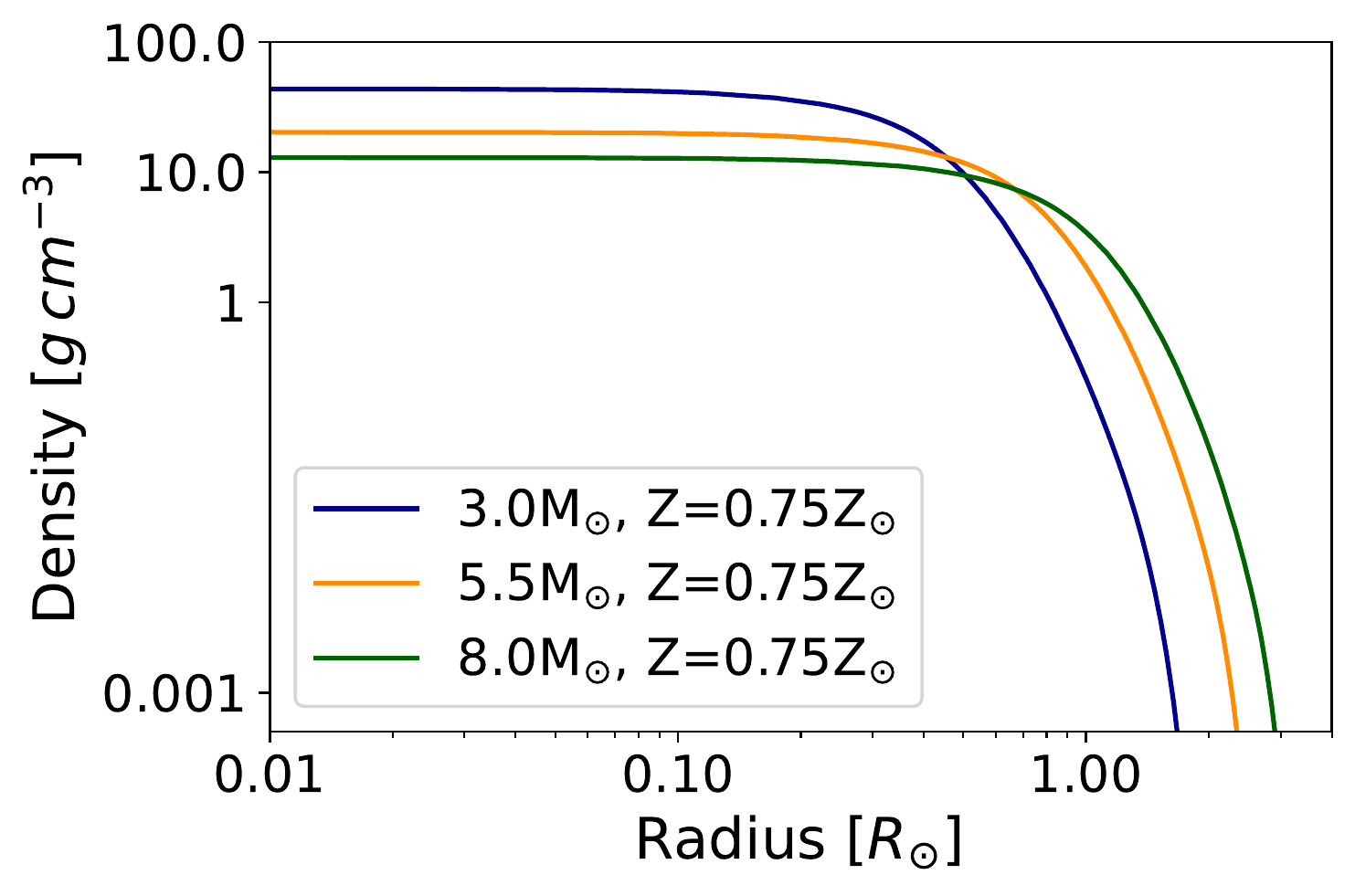}
\caption{The density profiles of our considered companion models. Different color represents different companion mass.}
\label{fig:cpmod}
\end{figure}

The companion metallicity is assumed to be identical to the supernova metallicity, $Z_{*}=0.75\, Z_{\rm \odot}$ \citep{2022ApJ...924...55G}. 
Figure~\ref{fig:cpmod} shows the density profiles of these three companion models. We also consider 6 different binary separations, varying from 3 to 8 $R_*$, where some of them slightly exceed the binary separation constraint. 
We set the explosion energy, $E_{\rm SN} = 7.9\times 10^{50}$~erg, based on the derived explosion energy of SN 2020oi \citep{2022ApJ...924...55G}.
These setups lead to 18 models in total. We also generate 4 more models with different explosion energy, 0.2, 0.5, 2, 5 $\times E_{\rm SN}$ to find the explosion energy dependence (Section \ref{sec:Edependence}). The total 22 simulations are summarized in Table~\ref{tab:modelsummary}.

\subsection{Initial setup}
We conduct axisymmetric 2D hydrodynamics simulations in \texttt{FLASH} to study the occurrence of supernova-companion interaction. The Helmholtz equation of state \citep{2000ApJS..126..501T} is used by interpolating a Helmholtz free energy table. The self-gravity is carried out using the improved multipole Poisson solver with the maximum multipole moment $l_{\rm max}=80$ \citep{2013ApJ...778..181C}.
The simulation domain of our \texttt{FLASH} simulations is a cylinder with 15$R_*$ in the $+$ radial (r) direction and each 15$R_*$ in the $\pm$ axial (z) directions in the cylindrical coordinates. The companion is located at the coordinate center, and the supernova ejecta is located one separation from the companion in the +z direction. The maximum refinement level is set to 9, corresponding to an effective uniform resolution of 4096 $\times$ 8192. The inner boundary condition is set to be "reflect" on the z-axis for the r-direction, and the outer boundary conditions are "outflow". We do not consider the orbital motion and spin of the companion because both of their velocities are much smaller than the explosion velocity ($\rm \sim 10^4 \, km \, s^{-1}$). We do not consider the nuclear reaction in the \texttt{FLASH} domain as well. The composition of the companion is simplified to be hydrogen ($\rm ^1H$), helium ($\rm ^4He$), carbon ($\rm ^{12}C$), and oxygen ($\rm ^{16}O$) only.

\subsection{Supernova-companion Interaction Simulations}
After generating the \texttt{MESA} companion models, we map the spherically symmetric \texttt{MESA} models into the 2D cylindrical grids in \texttt{FLASH}. We relax the companion models by damping the velocity magnitude in $\sim$300 dynamical timescales to reduce the grid artifacts on model interpolation \citep{2008ApJ...672L..41R,2012ApJ...750..151P, 2022ApJ...933...38R}. Since we do not focus on the detailed core-collapse supernova explosion mechanism, the supernova is simplified to a Sedov-like explosion within a radius of 15 times the smallest cell space ($\Delta x$), $R_{\rm SN} = 15\Delta x$. This is thought to be small enough to be considered as a point source compared to the companion size and avoid grid effects on the explosion \citep{2012ApJ...750..151P}. Note that the value of the smallest cell space in our setup depends on the size of the companion model. 

We take the SN explosion conditions from the suggested observed values of the SN 2020oi in \cite{2022ApJ...924...55G} (see Table \ref{tab:SN2020oi}). The explosion energy ($E_{\rm SN}$) and the ejecta mass ($M_{\rm ej}$) are derived from the bolometric fitting applying the Khatami \& Kasen model \citep{2019ApJ...878...56K}, which is based on the Arnett model \citep{1982ApJ...253..785A}. The Khatami \& Kasen model modifies the assumption of the uniform internal energy distribution as light propagates through the optically-thick ejecta materials. The average ejecta velocity ($v_{\rm SN}$) is determined by the Si line spectroscopically.

The density profile of the explosion $\rho(r)$ is assumed to be uniform within an explosion radius $R_{\rm SN}$,
\begin{equation}
\label{eq:densprofile}
    \rho(r) = M_{\rm ej} / \left( \frac{4 \pi R_{\rm SN}^3}{3} \right),
\end{equation}
and the velocity distribution $v(r)$ is assumed to be linearly proportional to radius,
\begin{equation}
\label{eq:vprofile}
    v(r) =  f_{\rm v} \cdot (\alpha \,  v_{\rm SN}) \, \left( \frac{r}{R_{\rm SN}} \right), 
\end{equation}
where $\alpha$ is a geometric factor in ensuring the total kinetic energy equal to the explosion energy. 
When the hydrodynamics system starts to evolve, a part of the internal energy is converted into kinetic energy in the first few minutes; this causes the total kinetic energy of the explosion at later times to be slightly larger than the expected observed value. Therefore, we multiply an extra factor $f_{\rm v}$ in the velocity profile in Equation \ref{eq:vprofile} to adjust the final kinetic energy.
$f_{\rm v} = 0.86$ is used in our simulations.
We assume the ratio of the supernova internal to kinetic energy $E_{\rm int}/E_{\rm tot}$ = 0.25 \citep{2015ApJ...806...27P}.
We don't consider the gravitational force from the remnant neutron star (NS) or black hole (BH) for simplification. 
The composition of the supernova ejecta is approximated to be 70\% helium ($\rm ^{4}He$), 10\% carbon ($\rm ^{12}C$), 10\% oxygen ($\rm ^{16}O$), and 10\% nickel ($\rm ^{56}Ni$), based on the reported early time abundance patterns and the nickel mass of SN 2020oi \citep{2022ApJ...924...55G}.

Due to the small time step in hydrodynamics simulation, doing the post-impact companion evolution with \texttt{FLASH} is computationally expensive. In addition, we do not consider nuclear burning in {\tt FLASH} simulations. 
Therefore, we conduct the post-impact simulations of the surviving companions in {\tt MESA}. 
Since the companion returns back to a relatively stable and spherically symmetric state in $\sim$100-200 hours after the SN explosion, we choose a snapshot of each \texttt{FLASH} model at t=150 hours, where the entropy profile is stabilized near the surface, as the input model of \texttt{MESA} to measure the entropy changes ($\Delta s$) from the initial model.

\subsection{Post-impact Companion Models}
\label{sec:nr_postevo}

To mimic the SN-impacted companion in {\tt MESA}, we artificially strip the companion mass and inject additional energy in the envelope of the original {\tt MESA} companion model used for the {\tt FLASH} simulations. 
The amount of mass loss is determined by the difference between the final mass of a surviving companion in {\tt FLASH} and the initial mass. 
We follow the heating process described in \cite{2019ApJ...887...68B} and \cite{2022ApJ...933...38R}, in which the heating rates are measured by $\dot{\epsilon} = T \Delta s / \Delta t$, where T is the local temperature, and $\Delta t$ is the heating timescale, which we set $\Delta t = $ 3 days. 
Mass loss is performed by the relaxation process \texttt{relax\_mass\_scale} of \texttt{MESA}. This process relaxes the mass of a star to the assigned mass without changing the composition.
Since the amount of mass loss is small, $\Delta M_* / M_{\rm *} \lesssim 1\%$ in all of our models (see Table \ref{tab:modelsummary}), the composition changes are negligible.
In addition, we also discuss and compare our heating method with the approach proposed by \cite{2021MNRAS.505.2485O} in Section \ref{sec:Ogata} as the post-impact evolution resulting from each method exhibits some discrepancies that could affect the prediction of the detectable period of the surviving companion.

\startlongtable
\begin{deluxetable*}{cccccccc}
\tablewidth{0pt}
\tablenum{2}
\tablecaption{Summary of the models.}
\tablehead{
Model &
$M_{*}$ [$ M_{\odot}$] &
$M_{*,f}$ [$ M_{\odot}$] &
$\Delta M_*$/$M_{*}$ [\%] &
$R_*$ [$\rm R_{\odot}$] &
$A$ [$R_*$]& 
$E_{\rm SN}$ [erg] &
$v_{\rm kick}$ [km $\rm s^{-1}$]
}
\startdata
M3.0A3  &  3.0  &  2.9672  &  1.093  &  1.93  &  3.0  &  7.9e+50  &  21.034\\
M3.0A4  &  3.0  &  2.9869  &  0.436  &  1.93  &  4.0  &  7.9e+50  &  11.204\\
M3.0A5  &  3.0  &  2.9939  &  0.204  &  1.93  &  5.0  &  7.9e+50  &  8.318\\
M3.0A6  &  3.0  &  2.9962  &  0.126  &  1.93  &  6.0  &  7.9e+50  &  5.012\\
M3.0A7  &  3.0  &  2.9972  &  0.092  &  1.93  &  7.0  &  7.9e+50  &  4.272\\
M3.0A8  &  3.0  &  2.9977  &  0.076  &  1.93  &  8.0  &  7.9e+50  &  3.353\\
M5.5A3 (1.0E) &  5.5  &  5.4708  &  0.531  &  2.71  &  3.0  &  7.9e+50  &  13.440\\
M5.5A4  &  5.5  &  5.4914  &  0.156  &  2.71  &  4.0  &  7.9e+50  &  6.819\\
M5.5A5  &  5.5  &  5.4963  &  0.068  &  2.71  &  5.0  &  7.9e+50  &  3.996\\
M5.5A6  &  5.5  &  5.4971  &  0.053  &  2.71  &  6.0  &  7.9e+50  &  2.794\\
M5.5A7  &  5.5  &  5.4982  &  0.033  &  2.71  &  7.0  &  7.9e+50  &  2.148\\
M5.5A8  &  5.5  &  5.4981  &  0.034  &  2.71  &  8.0  &  7.9e+50  &  1.817\\
M5.5A3 (0.2E) &  5.5  &  5.4957  &  0.078  &  2.71  &  3.0  &  1.6e+50  &  3.672\\
M5.5A3 (0.5E) &  5.5  &  5.4855  &  0.264  &  2.71  &  3.0  &  4.0e+50  &  8.749\\
M5.5A3 (2.0E) &  5.5  &  5.4413  &  1.066  &  2.71  &  3.0  &  1.58e+51 &  20.346\\
M5.5A3 (5.0E) &  5.5  &  5.3766  &  2.244  &  2.71  &  3.0  &  3.95e+51 &  35.138\\
M8.0A3  &  8.0  &  7.9754  &  0.307  &  3.36  &  3.0  &  7.9e+50  &  9.534\\
M8.0A4  &  8.0  &  7.9913  &  0.109  &  3.36  &  4.0  &  7.9e+50  &  5.165\\
M8.0A5  &  8.0  &  7.9963  &  0.047  &  3.36  &  5.0  &  7.9e+50  &  2.959\\
M8.0A6  &  8.0  &  7.9967  &  0.041  &  3.36  &  6.0  &  7.9e+50  &  2.184\\
M8.0A7  &  8.0  &  7.9981  &  0.024  &  3.36  &  7.0  &  7.9e+50  &  1.443\\
M8.0A8  &  8.0  &  7.9985  &  0.019  &  3.36  &  8.0  &  7.9e+50  &  1.149
\enddata
\tablecomments{
$M_{*}$ is the pre-SN companion mass, $M_{*,f}$ is the post-SN companion mass, $\Delta M_*/M_{*}$ is the mass loss fraction,
$R_*$ is the initial companion radius, $A$ is the initial binary separation in the unit of companion radius, and $v_{\rm kick}$ is the kick velocity of the surviving companion.
}
\label{tab:modelsummary}
\end{deluxetable*}

\section{RESULTS}
\label{sec:results}

\begin{figure*}
\epsscale{1.0}
\plotone{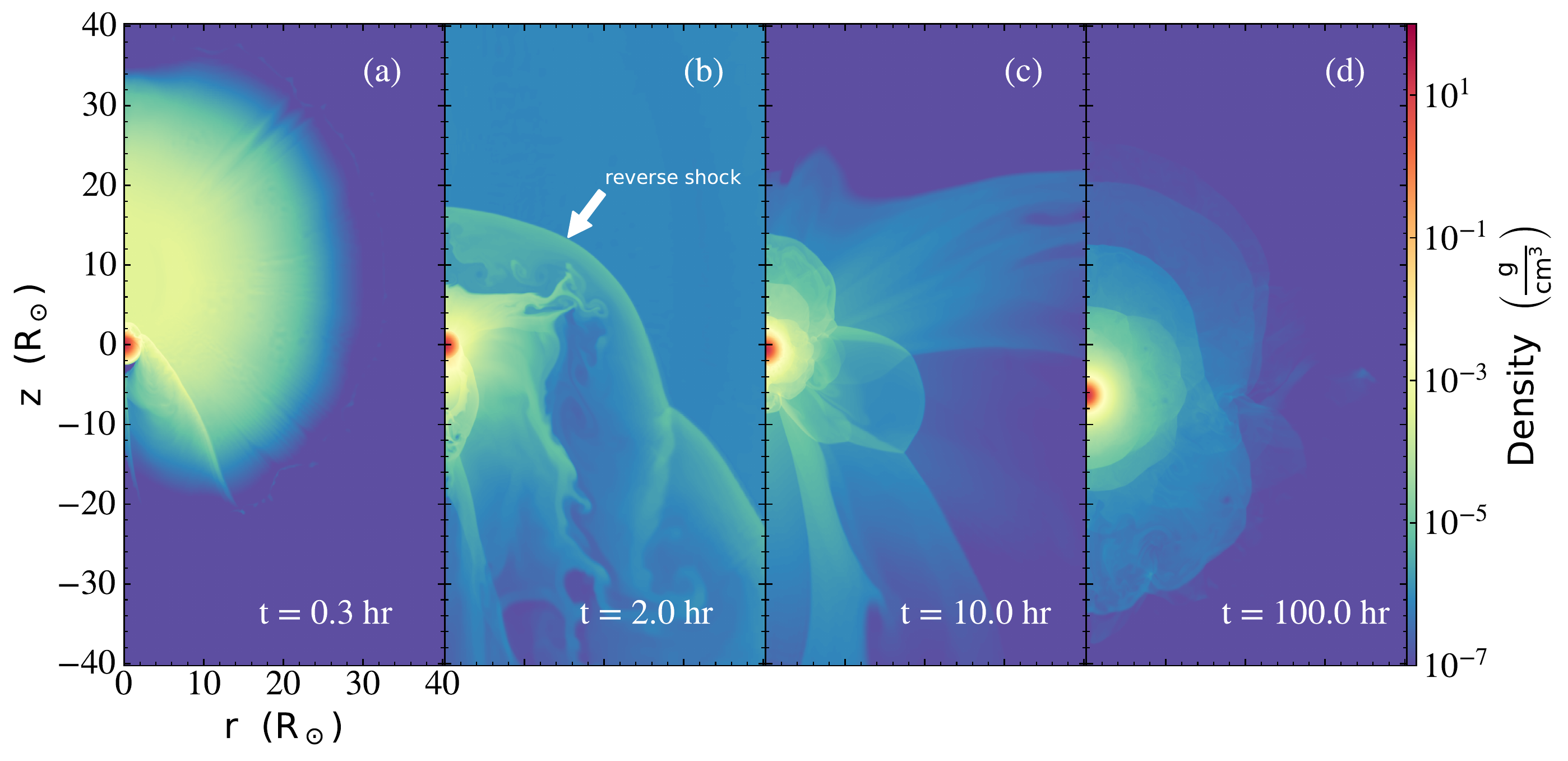}
\caption{Gas density distribution of model \texttt{m5.5a3e7.9} in the cylindrical coordinates at (a) t=0.3 hr (b) t=2.0 hr (c) t=10.0 hr (d) t=100.0 hr after the SN explosion.}
\label{fig:densslice}
\end{figure*}

\begin{figure}
\epsscale{1.0}
\plotone{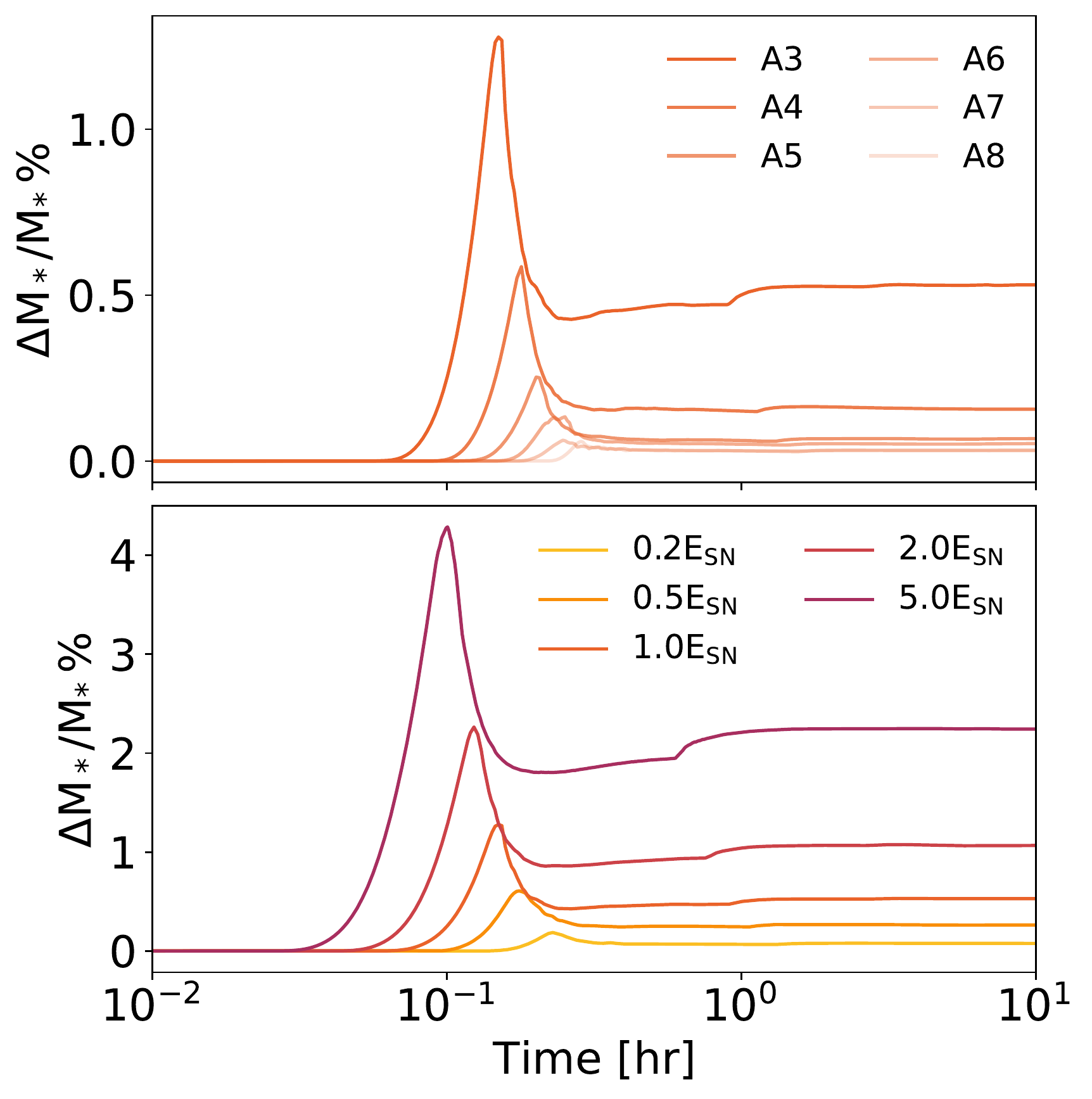}
\caption{Unbound mass fraction as functions of time for models with $M_* = 5.5 M_{\odot}$ and with different initial binary separations (upper panel) and explosion energies (lower panel). The upper panel uses the default explosion energy $E_{\rm SN}=7.9\times 10^{50}$ erg, while the lower panel assumes a binary separation $a=3R_*$.} 
\label{fig:unboundmass}
\end{figure}

\begin{figure}
\epsscale{1.0}
\plotone{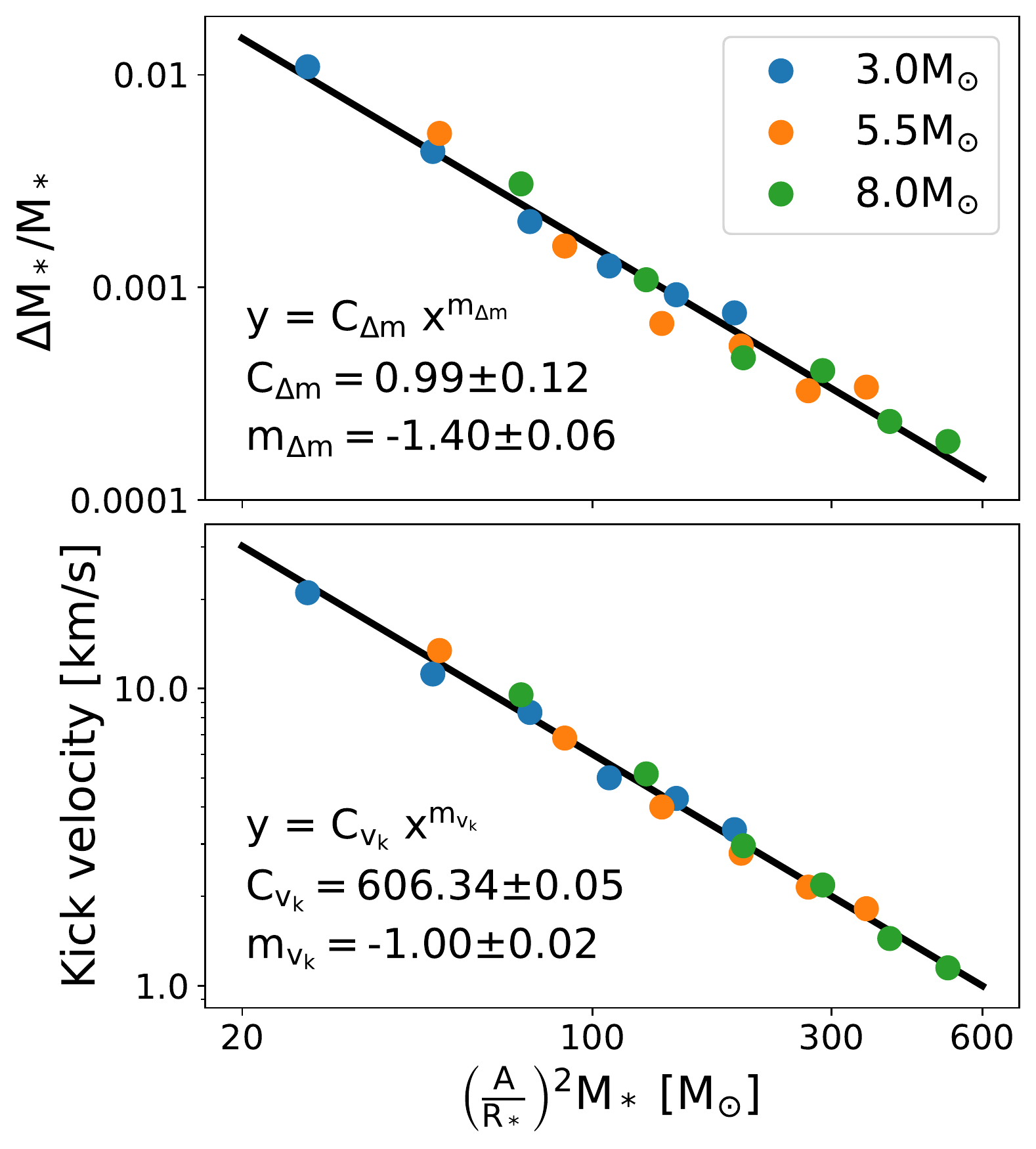}
\caption{Fraction of total unbound mass and kick velocity as functions of $(\frac{A}{R_*})^2 M_*$ with a fixed explosion energy $E_{\rm SN} = 7.9 \times 10^{50}$ erg. 
Different color indicates models with different companion mass.
The black line shows the best-fitted power-law relation as described in Equations~\ref{eq:ma} and \ref{eq:va}.}
\label{fig:MandvofA}
\end{figure}

\begin{figure}
\epsscale{1.0}
\plotone{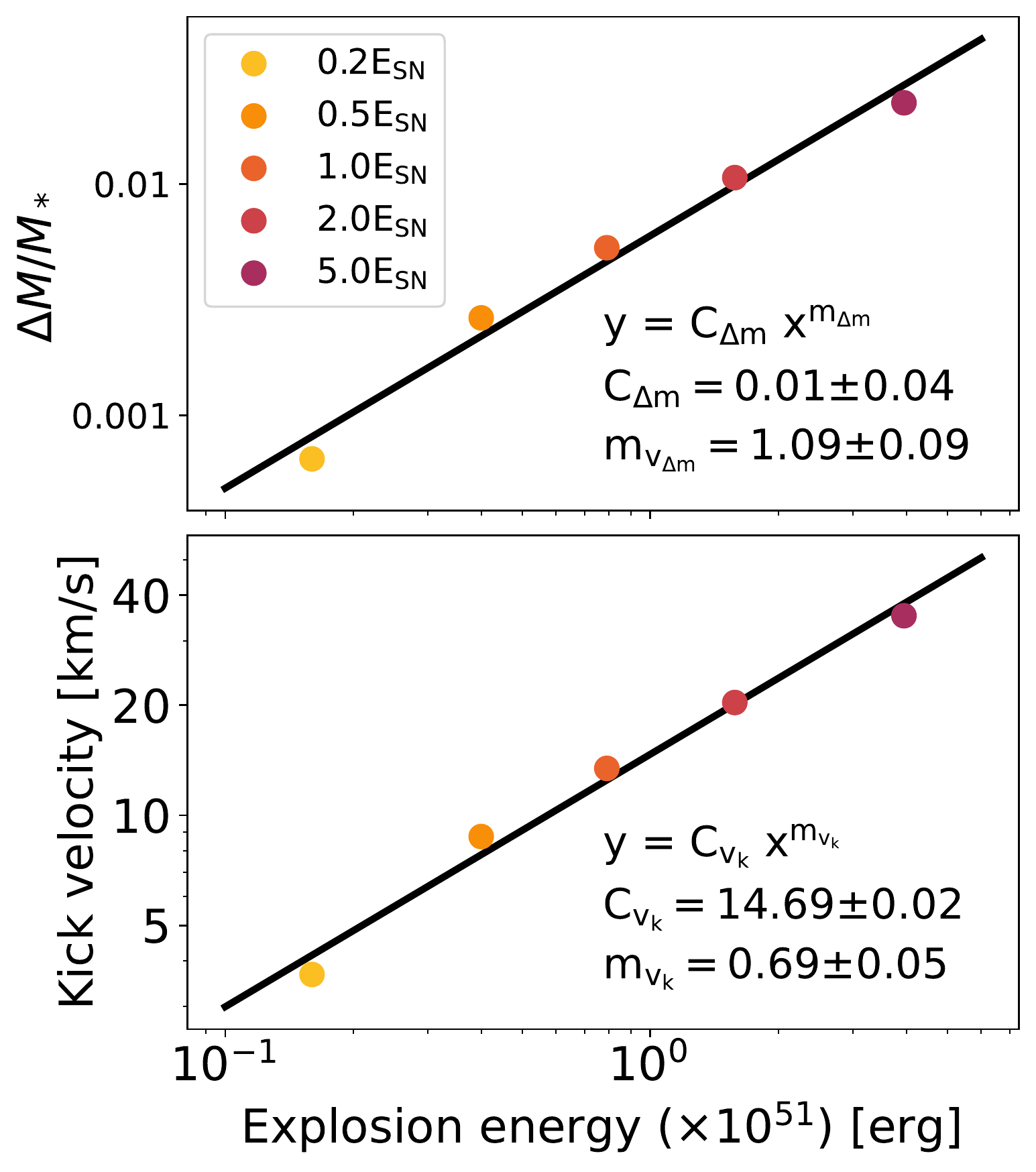}
\caption{Fraction of total unbound mass and kick velocity as functions of explosion energy with fixed companion mass $M_*$ = 5.5 $\rm M_{\odot}$ and binary separation $A$ = 3.0 $R_*$.
Different color indicates models with different explosion energy. Note that the $E_{\rm SN}$ in labels represents the default explosion energy, $E_{\rm SN}=7.9\times 10^{50}$ erg (see Table~\ref{tab:SN2020oi}).
The black lines show the best-fitted power-law relation as described in Equations~\ref{eq:M(E)} and \ref{eq:v(E)}.}
\label{fig:MandvofE}
\end{figure}

\begin{figure}
\epsscale{1.0}
\plotone{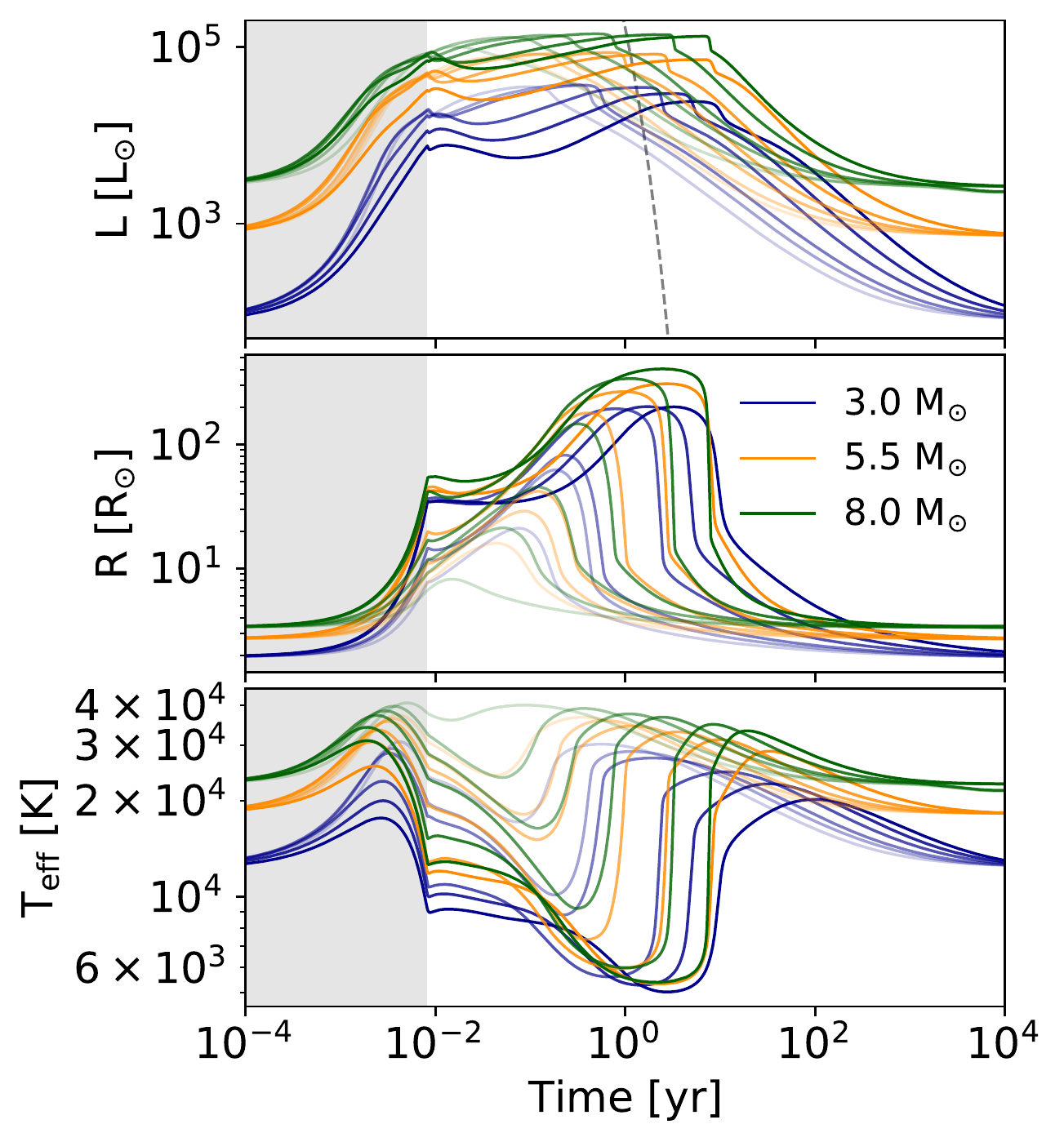}
\caption{The post-impact surviving companion evolution of bolometric luminosity, photosphere radius, and effective temperature for all of the models with explosion energy $E_{\rm SN}=7.9\times10^{50}$ erg. The colors represent different companion masses, and the line transparency is proportional to the separation. There are three different sets of companion mass, [3.0, 5.5, 8.0]$\times M_{\odot}$, and each companion mass has six different binary separations, ranging from 3, 4, 5, 6, 7, to 8$\times R_*$, in total 18 models. The gray area represents where the surviving companions are under heating. The heating timescale is 3 days. The dashed line in the upper panel represents the predicted lightcurve of SN 2020oi using the observed 40-65 days data in \cite{2022ApJ...924...55G} and fitted by the theoretical lightcurve model described in \cite{2015MNRAS.450.1295W}.}
\label{fig:post_all}
\end{figure}

\begin{figure*}
\epsscale{1.0}
\plotone{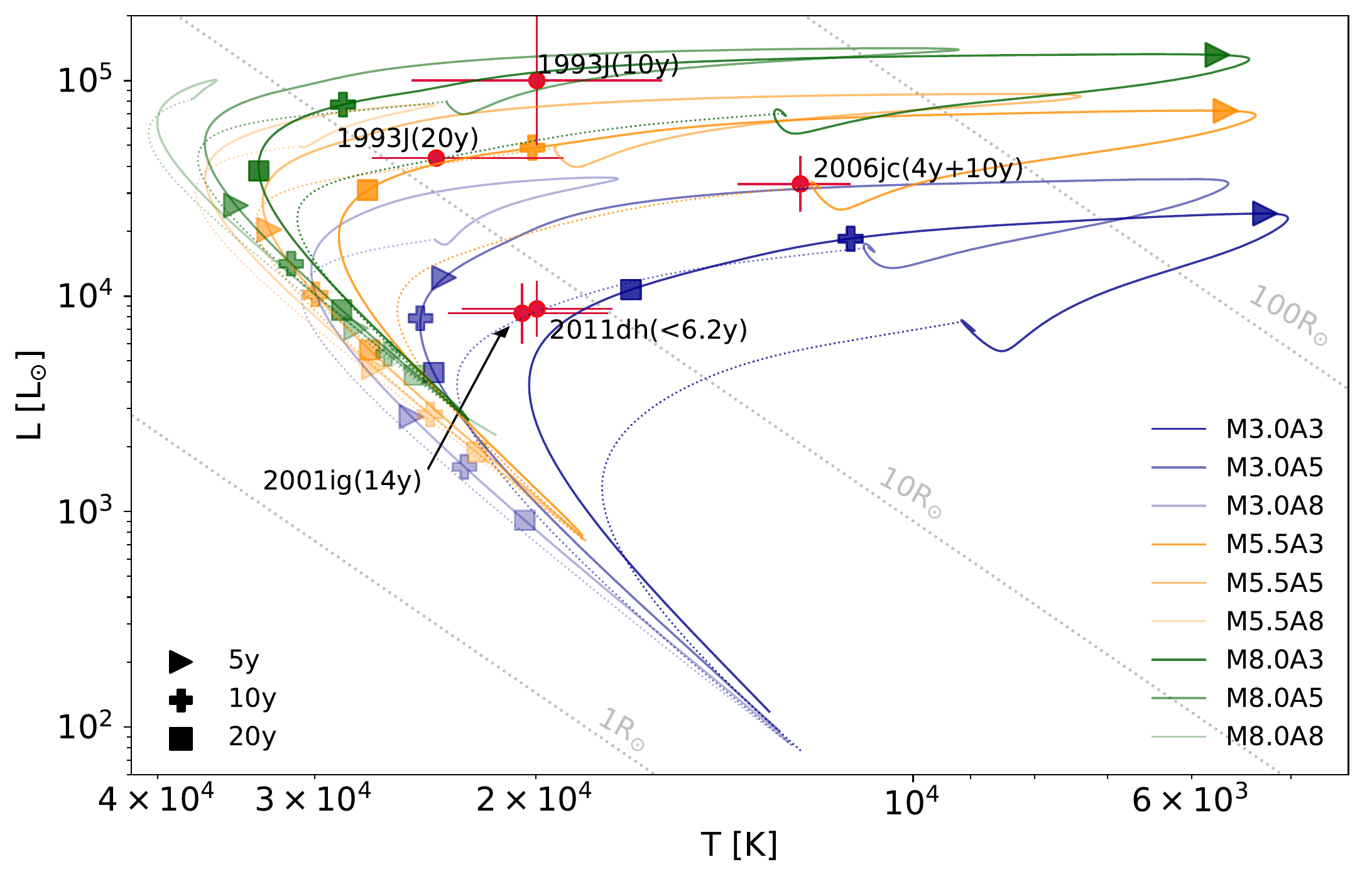}
\caption{Post-impact companion evolution on the H-R diagram. The colors represent different companion masses of the models. The transparency of lines represents the binary separation: the most opaque one is the closest one (A3 models), and the most transparent one is the farthest one (A8 models). Other models (A4-A7) are distributed in between but are not all plotted in the figure for clarity. The symbols mark the locations of surviving companions on the H-R diagram at times of 5, 10, and 20 years after the SN explosion. The red dots with error bars show the SED fitting results of the possible SE SNe companions to date: SN 1993J(a) \citep{2004Natur.427..129M}, SN 1993(b) \citep{2014ApJ...790...17F}, SN 2001ig \citep{2018ApJ...856...83R}, SN 2006jc \citep{2020MNRAS.491.6000S}, and SN 2011dh \citep{2019ApJ...883...86M}.}
\label{fig:hr}
\end{figure*}

\subsection{Supernova-companion Interaction Simulations}

Figure \ref{fig:densslice} shows the gas density distribution of model $\rm M5.5A3$ at different times. 
The overall phenomenon is similar to what has been described in the previous supernova-companion interaction studies \citep{2012ApJ...750..151P, 2015A&A...584A..11L, 2018ApJ...864..119H}.
First, the SN ejecta strips the companion's envelope and forms a bow shock at the companion's surface toward the ejecta (see Figure~\ref{fig:densslice}a). The ram pressure of SN ejecta gives a kick to the companion's center of mass to the $-z$ direction; all of the companion models get a kick velocity insufficient to set the companion unbound (see Table \ref{tab:modelsummary}).
Note that the kick velocity defined here is the velocity gain of the companion due to SN impact which is different from the kick velocity of a neutron star due to the asymmetry explosion of a CCSN.
In addition, the listed kick velocity values are upper limit since we don't consider the gravity from NS and the mass changes of the system before and after the SN explosion.
The companion is also compressed and heated by the SN ejecta. 
Once the SN shock passes through the companion, a reverse shock of SN ejecta can be seen in Figure~\ref{fig:densslice}b.
The energy deposition from the SN ejecta makes the companion envelope inflate, and the compression and fluid instability make the companion oscillate and form layers of radial shocks (Figure \ref{fig:densslice}c). After $\sim 100$ hours, the companion starts to resume equilibrium (Figure \ref{fig:densslice}d).

Figure \ref{fig:unboundmass} presents the systematic unbound mass fraction versus time of the 5.5 $\rm M_{\odot}$ cases. The upper panel shows the models varying separation from 3 to 8 $\rm \times R_*$ with fixed explosion energy, $E_{\rm SN}$. 
The lower panel shows models with different explosion energy but with a fixed binary separation, 3$R_*$.
All models follow a similar trend: The first peak with high unbound mass is due to shock compression.
Once the companion is released, some of the unbound mass resumes bound again.
Afterward, the curve first goes up gently, then forms a steeper slope at $\sim$1 hour, and finally becomes stable. 
All of our models reach stable companion mass within 10 hours, so we record the mass at 10 hours as the final companion mass. The initial mass, final mass, and unbound mass fraction of the companion models are listed in Table \ref{tab:modelsummary}.

\subsection{Binary Separation Dependence}
\label{sec:Adependence}

Figure~\ref{fig:MandvofA} shows the fraction of final unbound mass and kick velocity as functions of initial binary separations. Power-law relations on both unbound mass and kick velocity are consistent with what has been discussed in the past in \cite{2015A&A...584A..11L, 2018ApJ...864..119H, 2022ApJ...933...38R}, but in Figure~\ref{fig:MandvofA} we normalized the binary separation of a companion based on its radius and mass. With this normalization, universal power-law relations can be fitted by, 
\begin{equation}
\label{eq:ma}
    \frac{\Delta M_*}{M_{\rm *,i}} = (0.99 \pm 0.12) \times \left[ \left( \frac{A}{R_*} \right)^2 M_* \right] ^{-1.40 \pm 0.06},
\end{equation}
and
\begin{equation}
\label{eq:va}
    v_{\rm kick} = (606.34 \pm 0.05) \times \left[ \left( \frac{A}{R_*} \right)^2 M_* \right]^{-1.00 \pm 0.02}  [{\rm km \, s}^{-1}].
\end{equation}
The normalization factors $(A/R_*)^2 M_*$ can be understood by the combination of the inverse of the impact cross-section and the companion mass. The larger the $(A/R_*)^2$, the smaller the impact cross-section and, therefore, the lower the final unbound mass. In addition, a higher companion mass $M_*$ requires larger incident energy to unbound the same fraction of mass and also has a lower kick velocity when it receives the same amount of ram pressure and energy. 

\subsection{Explosion Energy Dependence}
\label{sec:Edependence}

To examine the energy dependence, we perform additional five models with fixed companion mass $M_* = 5.5 M_{\odot}$, fixed separation $A=3\times R_*$, but set the explosion energy ($E_{\rm SN}$) to 0.2, 0.5, 1, 2, 5 $\times E_{\rm SN}$, where $E_{\rm SN}=7.9\times 10^{50}$~erg (see Table \ref{tab:modelsummary}).
The lower panel of Figure \ref{fig:unboundmass} shows the unbound mass evolution of these models.
The fraction of final unbound mass and kick velocity can be fitted by,
\begin{equation}
\label{eq:M(E)}
    \frac{\Delta M_*}{M_{\rm *,i}} = (0.01 \pm 0.04) \times \left( \frac{E_{\rm SN}}{10^{51} \, \rm erg} \right)^{1.09} \pm {0.09},
\end{equation}
and
\begin{equation}
\label{eq:v(E)}
    v_{\rm kick} = (14.69 \pm 0.02) \times \left( \frac{E_{\rm SN}}{10^{51} \, \rm erg} \right)^{0.69 \pm 0.05}  [{\rm km \, s}^{-1}],
\end{equation}
respectively {and shown in Figure~\ref{fig:MandvofE}.} 

The unbound mass is linearly proportional to the explosion energy, which is consistent with previous works \citep{2008A&A...489..943P, 2015A&A...584A..11L, 2022ApJ...933...38R}. The kick velocity is close to but not accurately proportional to $\sqrt{E_{\rm SN}}$ because a portion of the supernova energy is used to strip and ablate the companion.

\subsection{Post-impact Companion Evolution}
\label{sec:post_evo}

To investigate the long-term post-impact evolution of surviving companions, we convert the 2D hydrodynamics models back into {\tt MESA}, using the heating process described in Section \ref{sec:nr_postevo}.
Figure \ref{fig:post_all} shows the post-impact surviving companion evolution of 18 models with different companion masses and binary separations (the full post-impact evolution data of these 18 models are available on Zenodo: \dataset[doi: 10.5281/zenodo.7777107]{https://doi.org/10.5281/zenodo.7777107}.)
When the heating process ends on day 3 after the SN explosion, the deposited energy slowly radiates away, making the peak bolometric luminosity and photosphere radius of all our models grow 1-2 orders larger than the original state.
The luminosity reaches the peak value from months to a few years, depending on their initial binary separation and local thermal timescale. In principle, shorter binary separations lead to stronger SN impact and, therefore, deeper energy deposition \cite{2012ApJ...760...21P, 2022ApJ...933...38R}.
Once the deposited energy fully radiates away, the surviving companions start to contract by releasing their gravitational energy. The effective temperature also increases due to this contraction. After $\sim 10^4$ years, all surviving companions eventually resume their original states before the SN explosion since the amount of mass loss is negligible. 

It should be noted that the effects from the remnant NS or BH are not considered in this analysis. 
If the binary system remains bound after the SN explosion, the expansion of the surviving companion would be limited by the resulting Roche-lobe radius. Roche-lobe overflow or common envelope might be formed in this kind of system.
However, \citet{2021MNRAS.505.2485O} conducted binary population synthesis simulations and discussed the possible outcomes of the post-SN system.
They found that $>90 \%$ of their models evolve to bound systems but have wide binary separations and no binary interaction.
Therefore, the companion radius and luminosity in our simulations are upper limits but applicable to most cases.

Figure \ref{fig:hr} shows the evolutionary tracks of our models on the H-R diagram. Different transparency of curves represents the different initial binary separations of the model.
Each model is composed of two line segments: a dotted line and a solid line. The evolutionary track starts from the lower luminosity region of the dotted line and becomes brighter in time, which represents the heating process by the SN impact. 
After the SN heating, the evolutionary track is shown by the solid line and then continues evolving to the upper right corner of the HR diagram. 
Note that the dotted curves drop a bit after the junction with solid lines; this can be seen in Figure \ref{fig:post_all} as well, where the luminosity curves drop slightly right after the heating process while the photosphere radii remain flat. 

In general, more massive companions have brighter evolution, and closer binary separations lead to more vigorous variations due to stronger SN impact. A stronger SN impact leads to a bigger companion inflation and a deeper energy deposition, and therefore has a cooler effective temperature and takes a longer time to resume its original state, but the value of peak luminosity is less sensitive to the initial binary separation. 
Since we have explored a wide range of companion masses and initial binary separations, our results, in principle, could be extended to other SN~2020oi-like systems and other SE SNe with a similar companion mass as well.

\begin{figure*}
\epsscale{1.0}
\plotone{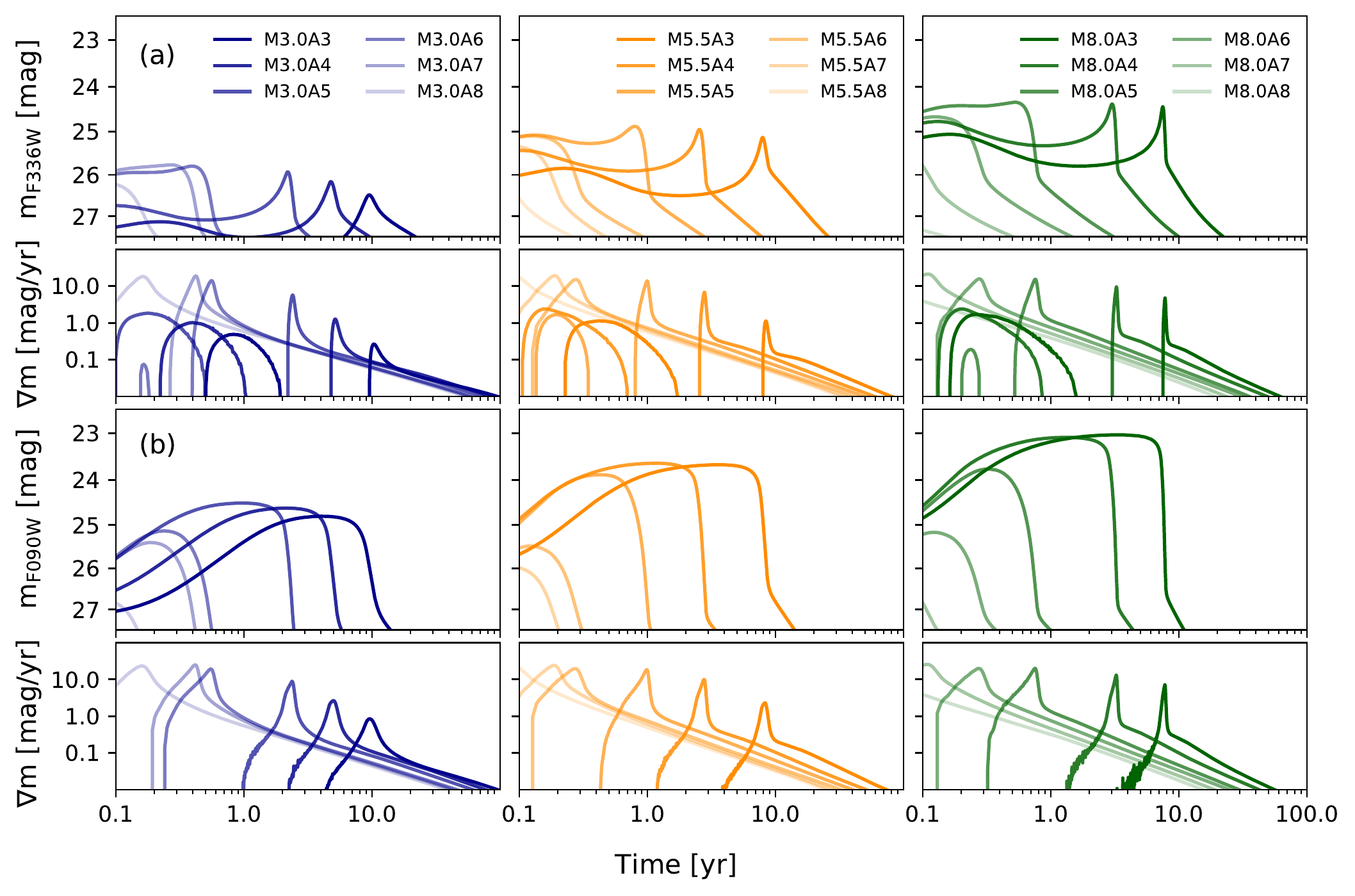}
\caption{Estimated apparent magnitude evolution of the surviving companion of SN 2020oi with HST filter F336W (upper panels) and JWST filter F090W (lower panels). The first and third rows show the magnitude evolution with different models, and the second and fourth rows represent the evolution of the magnitude gradient. The color represents different companion masses, and the line transparency is proportional to the initial binary separation. The combined reddening is assumed to be $E(B-V)=0.133$ mag, and the distance applied is 17.1 Mpc \citep{2022ApJ...924...55G}.}
\label{fig:sn2020oi}
\end{figure*}

\section{DISCUSSION}
\label{sec:discussion}

In this section, we provide several observational signatures of the possible surviving companion of SN~2020oi and discuss the observability of the surviving companion in the following decade. 
In addition, we discuss several surviving companion candidates that were reported in the literature and compare them with our models. 
Finally, we compare our heating method in this study with the method used in \citet{2021MNRAS.505.2485O} and show that the evolutionary timescale of surviving companion is sensitive to the detailed treatment of SN heating, but the overall trend remains the same.

\subsection{Observability of the surviving companion in SN 2020oi remnant}
\label{sec:obs_SN2020oi}

To search for the surviving companion of SN~2020oi, we further evaluate the apparent magnitude evolution with the filters\footnote{\href{http://svo2.cab.inta-csic.es/theory/fps/}{http://svo2.cab.inta-csic.es/theory/fps/}} of the Hubble Space Telescope (HST) and James Webb Space Telescope (JWST), using the Filter Profile Service of the Spanish Virtual Observatory (SVO).
Since surviving companions turn red a little bit ($T_{\rm eff} \sim 5000 - 12000$~K) in most of our models after the SN explosion (see Figure~\ref{fig:hr}), we choose HST filters F336W, F438W, and F555W which are sensitive in the bluer bands of optical light.
On the other hand, JWST has better sensitivity and spatial resolutions and can identify deeper objects than HST, but JWST's instruments focus more on infrared bands. We consider near-infrared (NIR) short-wavelength filters F070W, F090W, F115W, and F150W of JWST.
Galactic extinction is considered for each SVO filter using the extinction law by \citet{1999PASP..111...63F} with improvement by \citet{2005ApJ...619..931I} in the infrared. 
The distance of SN~2020oi is assumed to be $d = 17.1$ Mpc and the combined reddening of SN 2020oi is $E(B-V)=0.133$ mag, assuming $R_V$=3.1 \citep{2022ApJ...924...55G}.

Figure \ref{fig:sn2020oi} shows the apparent magnitude evolutionary tracks with the HST F336W and JWST F090W filters of the 18 fixed-energy models in Table~\ref{tab:modelsummary}. (Magnitude evolution with other filters are available on Zenodo: \dataset[doi: 10.5281/zenodo.7777107]{https://doi.org/10.5281/zenodo.7777107}.)
In general, the evolutionary tracks are similar to those of the luminosity evolution (Figure \ref{fig:post_all}), but the detailed shapes of the tracks in each filter are different due to the evolution of effective temperatures in each model.
It should be noted that before the SN explosion, the companion stars were MS A - B stars, which might be difficult to be detected by the HST, except for massive companions. After the SN impact, the surviving companion during the fast expansion phase ($3 -10$~yrs after the explosion) turns into an A, F, or G star, which could show significant brightening in optical and NIR bands. In Figure \ref{fig:sn2020oi}, we can see that most of our shorter-separation models have apparent magnitudes brighter than the typical short-exposure JWST NIRCam magnitude limit (26)\footnote{https://jwst-docs.stsci.edu/jwst-near-infrared-camera/nircam-performance/nircam-imaging-sensitivity/}. On the other hand, longer-separation models ($A>5 \rm R_*$) have less impact and brightening and, therefore, will not able to be detected by JWST or HST due to the fast decline in its magnitudes.      

Our results suggest that if the surviving companion's mass ranged within 3 to 8 solar masses and binary separation were within $5 R_*$, a surviving companion is expected to be detected by JWST and HST within roughly 10 years. In addition, the magnitude gradients could reach $> 1-10$~magnitude per year within 10 years after the SN explosion (see the second and the fourth rows in Figure~\ref{fig:sn2020oi}). Therefore, measuring the changes of magnitude would be another strong smoking gun evidence for the surviving companion.

However, one should note that the supernova remnant could still be bright and opaque after a few years of the explosion. The surviving companion can be detected only when the SN remnant becomes transparent and fainter than the surviving companion. 
\cite{2022ApJ...924...55G} provides the data of bolometric luminosity measured in the first 65 days since the explosion. Using the 40-65 days data and the late-time lightcurve model proposed by \cite{2015MNRAS.450.1295W}, we extrapolate the bolometric luminosity lightcurve up to 10 years after the SN explosion in Figure~\ref{fig:post_all}. The predicted lightcurve suggests that the SN 2020oi could be fainter than the surviving companion after two years.  
Combining these results, we recommend observers search for the surviving companion of SN~2020oi during the time window between 2023 to 2030. Furthermore, multiple observations to measure the changes in magnitudes and colors could provide another strong support for the identification of a surviving companion.

\subsection{Search for Surviving Companions in Stripped-envelope Supernova Remnants}
\label{sec: obs_SESNe}

Our simulation models could be applied to other SE SNe as well, as long as the binary configurations lay in the parameter space we considered. In Figure \ref{fig:hr}, we compare our models with four surviving companion candidates of SE SNe (red dots in Figure \ref{fig:hr}) that were reported in the literature. We discuss and compare some of the observational properties of these candidates with our models in this subsection.

\subsubsection{SN 1993J}
The binary companion of the Type~IIb SN~1993J was predicted and later confirmed by \citet{2004Natur.427..129M} to be a B2 star with $\log (T_{\rm eff}/\rm K)$ = 4.3 $\pm$ 0.1 and $\log (L/\rm L_{\odot})$ = 5.0 $\pm$ 0.3 (label 1993J(a) in Figure \ref{fig:hr}). These properties were obtained by comparing the spectral and photometric characteristics of the surviving companion to the standard single-star stellar evolution. The observed luminosity of the surviving companion is consistent with the predicted range of our M8.0 models but evolves slower than our predicted models.   
The temperature of the surviving companion falls within the range of our M5.5 and M8.0 models, except for those with large separations (A7 and A8 models).  

\cite{2014ApJ...790...17F} reexamined the surviving companion of SN1993J about 20 years after the explosion and found a hotter ($T_{\rm eff}$ = 19,000-27,000 K) and dimmer ($V_{\rm mag}$ = 22.9 mag) companion (label 1993J(b) in Figure \ref{fig:hr}). This trend is also consistent with the predicted evolutionary tracks of our models, which show the surviving companion moving toward the bottom left in the H-R diagram about 10 years after the SN explosion (i.e. model M8.0A3, although its temperature is higher than the observed temperature at the same age). In addition, the explosion energy of SN 1993J ($E_{\rm SN} \lesssim 10^{51}$ erg) and the amount of Nickel produced ($M_{\rm ^{56}Ni} = 0.07 \pm 0.01 \rm M_{\odot}$) are similar to those of SN 2020oi. Therefore, the surviving companion of SN 1993 might be similar to our model M8.0A3. 

Note that \cite{2004Natur.427..129M} estimate a companion mass of 22 $\rm M_\odot$  during the time of SN explosion based on the photometric and spectroscopic data ten years after the explosion. In addition, \cite{2009MNRAS.396.1699S} propose a binary evolution model with a similar initial mass range as that of \cite{2004Natur.427..129M}, but focusing on the effects of varying mass transfer efficiencies. Their model suggests that the companion at the time of explosion was very close to the MS stage.
However, these models don't consider the expansion of the surviving companion due to the SN impact and, therefore, might overestimate the companion mass and binary separation.

\subsubsection{SN 2001ig}
The surviving companion of Type~IIb SN 2001ig was detected by \cite{2018ApJ...856...83R} $\sim$ 14 years after SN explosion. The companion was identified as a B-type MS star with $T_{\rm eff}$=19,000-22,000 K and $\log(L/\rm L_{\odot})$=3.92$\pm$0.14 by fitting the UV photometry data to standard MS models.
Their best-fit binary evolution model from \texttt{BPASS} includes a MS companion with an initial mass of 9 $\rm M_{\odot}$ and an age near the terminal age main sequence (TAMS) when the primary exploded.
Compared with our models, the detected companion has a temperature closer to our longer-separation models ($\rm A5-A8$), which return to their original state earlier than the shorter-separation models.
Interestingly, at $\sim$ 3 years after SN, \citet{2006MNRAS.369L..32R} reported a possible detection of the companion, which was more like a supergiant with a spectral type ranging from late-B to late-F star. 
\cite{2018ApJ...856...83R} have discussed this discrepancy between the two epochs and ruled out a coincident nebula or a newly formed hot circumstellar medium (CSM) obscuring the companion.
Considering the SN impact scenario in our simulations, we propose another explanation that the companion at $\sim$ 3 years after SN was an inflated MS star and quickly returned to its original state as shown at $\sim$ 14 years after SN.

\subsubsection{SN 2006jc}
\cite{2016ApJ...833..128M} presented observations of Type~Ibn SN 2006jc at 2 and 4 years after the SN explosion, but only the latter observation detected a possible surviving companion.
\cite{2020MNRAS.491.6000S} confirmed the existence of the surviving companion using the 10-year-after-SN epoch and combined the data with the 4-year-after-SN data, as the magnitudes of the two epochs are comparable. By comparing the spectra of the combined data to supergiant models, they determined the companion to have a $\log(T_{\rm eff}/\rm K) = 4.09^{+0.05}_{-0.04}$ and $\log(L/\rm L_{\odot})$ = 4.52$\pm$0.13, corresponding to an evolved companion with $M_* \lesssim 12\, \rm M_{\odot}$. The \texttt{BPASS} models also suggest that the primary star had a similar size, but binary interaction was not considered in their analysis.
However, the estimated mass of the primary indicates that the system must have experienced binary interaction because the mass of the primary is insufficient to induce the stellar winds strong enough to strip all of its H~envelope.

In Figure~\ref{fig:hr}, we've shown that our models with less massive companions and encountered SN impact (M3.0A3) can also meet the observed properties of the surviving companion of SN~2006jc on the H-R diagram, e.g. $\rm M3.0$ models. In addition, since they combined the 4- and 10-year-after-SN observations, we cannot directly compare their observation with our models where the post-impact surviving companions vary rapidly within this time range.
\cite{2020MNRAS.491.6000S} also proposed a low-mass companion scenario based on their Eject-Companion-Interaction simulation. They suggest a 4 $\rm M_\odot$ companion with a binary separation $A=22.8 R_*$, which can also explain the observed properties. In contrast to their simulation, our simulations prefer shorter binary separation models while considering a similar mass companion. This difference might be due to the shallower energy deposition in our models and therefore evolve faster than the simulation in \cite{2020MNRAS.491.6000S} (and also in \citealt{2021MNRAS.505.2485O}). 
We will discuss this difference in more detail in Section~\ref{sec:Ogata}.  
Later epoch observations would place better constrain on the mass and evolution phase of the surviving companion of SN~2006jc.

\subsubsection{SN 2011dh}
SN~2011dh in M51 is a Type~IIb SN with pre-explosion HST observations.
The photometry data of the progenitor system is consistent with a binary model conducted by \cite{2013ApJ...762...74B}, which has the final progenitor and companion mass of 4 and 12 $\rm M_\odot$ and a final semi-major axis $\sim$ 900 $\rm R_{\odot}$ at the time of explosion. A blue point source at the location of SN~2011dh is observed by \cite{2014ApJ...793L..22F}, using near-UV HST observations at 1161 days after the explosion. \cite{2014ApJ...793L..22F} claim that part of the observed flux should come from the surviving companion (but disagreed by \citealt{2015MNRAS.454.2580M}). 

\cite{2019ApJ...883...86M} used the late-time HST observations from 1.8 to 6.2 years after the explosion and concluded that a B-type surviving companion with $\log(L/\rm L_{\odot})=3.94\pm0.13$ and $\log(T_{\rm eff}/\rm K) = 4.30\pm0.06$ could explain the late-time photometric lightcurve but cannot rule out the possibility of a slow rotating magnetar or a light echo originating from dust. 

If the observed point source is indeed a surviving companion, the late-time HST observations \citep{2019ApJ...883...86M} show a similar bolometric luminosity and effective temperature to the corresponding values in our model M3.0A5 (see Figure~\ref{fig:hr}), but the proposed companion mass (9-10 $\rm M_\odot$ by \cite{2019ApJ...883...86M} and 12 $\rm M_\odot$ by \cite{2013ApJ...762...74B})
as well as the final separation are much higher than our simulated values. We note that both binary models presented in \cite{2013ApJ...762...74B} and \cite{2019ApJ...883...86M} match the luminosity and effective temperature of their binary evolution model to the observed candidate. However, neither of these models takes into account the effect of SN impact on the surviving companion.

\begin{figure}
\epsscale{1.0}
\plotone{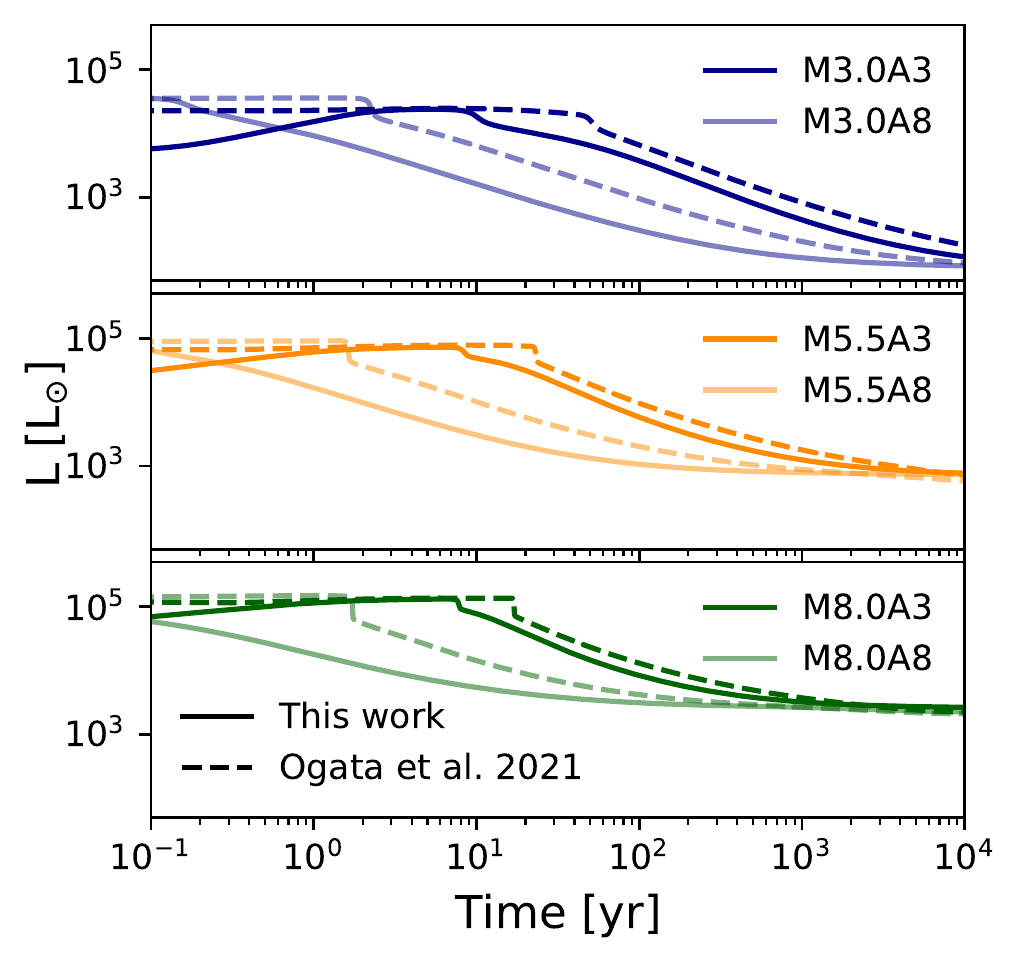}
\caption{The luminosity evolution of the post-impact companions in comparison with \citet{2021MNRAS.505.2485O}. The solid lines apply the heating procedure in this paper, and the dashed lines apply the heating procedure of \citet{2021MNRAS.505.2485O}.}
\label{fig:ogata}
\end{figure}

\subsection{Influence of the Heating Process -- Comparison with Ogata et al. (2021)}
\label{sec:Ogata}

In order to search for surviving companions of SE~SNe, the timing of the peak luminosity during the expansion phase is one crucial indicator of the progenitor properties (see Figure~\ref{fig:post_all}). We have noticed that although our predicted surviving companions are qualitatively similar to what have been predicted in \cite{2021MNRAS.505.2485O} but quantitatively different, especially the timing at the peak luminosity of surviving companions.  
This difference could be understood by the difference in supernova heating procedures since the expansion speed is determined by the local thermal timescale and sensitive to the depth of SN energy deposition. 

To examine the impact on the heating procedures, we ran 6 extra post-impact companion simulations using the heating formula described in \cite{2021MNRAS.505.2485O} to compare their heating method with ours.
Figure~\ref{fig:ogata} shows the comparison of bolometric luminosity evolution of models M3.0A3, M3.0A8, M5.5A3, M5.5A8, M8.0A3, and M8.0A8 using both heating methods. 
We find that simulations with our heating method show a faster decline than the corresponding simulations using \cite{2018ApJ...864..119H}'s method. The difference in the peak luminosity time could be as large as one order of magnitude, but the values of peak luminosity are comparable. 

It should be noted that \cite{2021MNRAS.505.2485O} use an universal heating formula based on their earlier work in \cite{2018ApJ...864..119H}, where
the heating formula is taken from a simple power-law fit of the entropy excess in their simulations. In \cite{2018ApJ...864..119H}, only large binary separations ($A > 20-60 R_*$) are considered, and the SN energy deposition mainly happens in the stellar surface due to weaker SN impact.
On the other hand, we calculate the entropy changes individually based on two-dimensional hydrodynamics simulations. Therefore, it seems that there could be some degeneracy in the companion masses, binary separations, and theoretical models. To better confirm a surviving companion, one can not only compare the location of surviving companion candidates on H-R diagram. Additional information, such as magnitude gradient, proper motion, rotation, or abundance abnormal, would be necessary.

\section{SUMMARY AND CONCLUSIONS}
\label{sec:sum&con}

Low-mass stripped-envelope supernovae are believed to be born in binary systems. One direct evidence to support this scenario is to find surviving companions after supernova explosions.
Based on the physical properties of the Type~Ic SN 2020oi reported in \citet{2022ApJ...924...55G}, we have conducted 18 axisymmetric {\tt FLASH} hydrodynamics simulations of the SN impact on a MS companion and performed subsequent post-impact evolution of surviving companions with the stellar evolution code {\tt MESA}. 
We find that the amount of stripped mass and kick velocity have power-law relations with their binary separations but the stripped mass and kick velocity are low compared to their original mass and orbital velocity and, therefore, it might be difficult to be detected, except in super-luminosity SNe with higher explosion energies.
This is due to the nature that in SE SNe, the binary companion is relatively more massive and compact than the SN ejecta.
However, the post-impact evolution shows significant brightening, inflation, and turning cool in the first few decades due to the energy deposition from SN. These transitions would be noticed by HST and JWST observations. The transitions also give a magnitude gradient in order of 1-10 magnitude per year, making it another smoking-gun evidence of a surviving companion.  
In addition, our results suggest that a surviving companion could be detected by HST and JWST starting from $\sim 2023 - 2030$. 

Furthermore, our simulated surviving companions are in order of magnitudes consistent with surviving companion candidates of SN~1993J, SN~2001ig, SN~2006jc, and SN~2011dh, but additional late-time observations and extra simulations with best-fit parameters are still necessary to confirm the progenitor systems of these SE~SNe. Finally, the comparison of our heating method with the method described in \cite{2021MNRAS.505.2485O} suggests the detailed heating procedure (or the depth of the SN energy deposition) are crucial on the post-impact evolution time scale, but the most bright luminosities are less sensitive with the heating methods.

This work is supported by the National Science and Technology Council of Taiwan through grant NSTC 111-2112-M-007-037, by the Center for Informatics and Computation in Astronomy (CICA) at National Tsing Hua University through a grant from the Ministry of Education of Taiwan. The simulations and data analysis have been carried out on the CICA Cluster at National Tsing Hua University and the Taiwania supercomputer at National Center for High-Performance Computing (NCHC). Analysis and visualization of simulation data were fulfilled with the analysis toolkit {\tt yt}.

\software{MESA r12115 \citep{2011ApJS..192....3P, 2013ApJS..208....4P, 2015ApJS..220...15P, 2018ApJS..234...34P, 2019ApJS..243...10P}, FLASH 4.5 \citep{2000ApJS..131..273F, 2008PhST..132a4046D}, yt \citep{2011ApJS..192....9T}, \href{https://github.com/jschwab/python-helmholtz}{python-helmholtz} \citep{2000ApJS..126..501T}, Matplotlib \citep{2007CSE.....9...90H}, NumPy \citep{2011CSE....13b..22V}, and SciPy \citep{2019zndo...3533894V}.}


\end{document}